\documentclass[12pt]{article}
\usepackage{graphicx}
\usepackage{amssymb}
\usepackage{amsfonts}
\usepackage{amsmath}
\usepackage[british]{babel}
\usepackage[utf8]{inputenc}

\usepackage[default]{jasa_harvard}
\usepackage{JABES_manu}

\usepackage{geometry}
\usepackage{setspace}
\title{The Gamma-count distribution in the analysis of experimental
  underdispersed data}

\author{Walmes Marques Zeviani$^1$\footnote{Corresponding author: walmes@ufpr.br, Dept Estat\'{\i}stica-UFPR, C.P. 19.081, Curitiba, PR, Brazil, 81.531-990},\\  
Paulo Justiniano {Ribeiro Jr}$^1$,\\
Wagner Hugo Bonat$^1$,\\
Silvia Emiko Shimakura$^1$,\\
Joel Augusto Muniz$^2$ \\
$^1$ {\small LEG/DEST - Paran\'{a} Federal University}\\
$^2$ {\small DEX/UFLA - Lavras Federal University}}

\date{}

\begin{document}

\maketitle

\newpage
\begin{center}
\textbf{Abstract}
\end{center}

Event counts are response variables with non-negative integer values
representing the number of times that an event occurs within a fixed
domain such as a time interval, a geographical area or a cell of a
contingency table. Analysis of counts by Gaussian regression models
ignores the discreteness, asymmetry and heterocedasticity and is
inefficient, providing unrealistic standard errors or possibily negative
predictions of the expected number of events. The Poisson regression is
the standard model for count data with underlying assumptions on the
generating process which may be implausible in many applications.
Statisticians have long recognized the limitation of imposing
equidispersion under the Poisson regression model. A typical situation
is when the conditional variance exceeds the conditional mean, in which
case models allowing for overdispersion are routinely used. Less
reported is the case of underdispersion with fewer modelling
alternatives and assessments available in the literature. One of such
alternatives, the Gamma-count model, is adopted here in the analysis of
an agronomic experiment designed to investigate the effect of levels of
defoliation on different phenological states upon the number of cotton
bolls. Results show improvements over the Poisson model and the
semiparametric quasi-Poisson model in capturing the observed variability
in the data. Estimating rather than assuming the underlying variance
process lead to important insights into the process.

\vspace*{.3in}


\newpage

\section{Introduction}\label{intro}
%% Regression models have been applied in the analysis of data gathered
%% from various areas of Science.  The linear (Gaussian) regression model
%% is without a doubt the most popular model among applied statistics
%% users.  Among a great variety of applications it is common to find
%% situations where the random variable of interest (response) is
%% presented in the form of counts. Often, counts are random variables
%% that assume non-negative integer values, representing the number of
%% times an event occurs within a fixed domain that can be continuous,
%% such as an interval of time or space, or discrete, such as the
%% evaluation of an individual or a census tract.

Regression models are deeply rooted in the analysis of agronomic
experiments and least squares methods associated to the linear
(Gaussian) model are widely adopted. On the other hand, response
variable in the form of counts are not uncommon. They may represent the
number of fruits produced by a tree, the number of units infected by a
disease, the number of insects on a particular plant structure, among
others. Counts are random variables that assume non-negative integer
values, representing the number of times an event occurs within a fixed
domain that can be continuous, such as an interval of time or space, or
discrete, such as the evaluation of an individual or a census tract.

Gaussian regression models for count data are not efficient, typically
producing inconsistent standard errors and even negative predictions for
the expected number of events \cite{King1989}. Gaussian linear model
ignores the discreteness, heterocedasticity, asymmetry and
non-negativeness, inherent features of count data. Impacts on the
results are greater when the sample size is small and the counts are
low.

Poisson regression became the standard model for count data, in
particular after the proposal of the unifying class of generalized
linear models \cite{Nelder1972} and the subsequent availability of
computational resources for model fitting.
%% The Poisson distribution naturally allows for asymmetric and
%% heterocedastic data, moreover, it has the non negative integer numbers
%% as its domain, making it a very appealing option to model count data.
The Poisson distribution is an appealing option to model count data
given its domain on the non negative integer numbers, moreover, it
naturally allows for asymmetry and heterocedasticity that are intrinsic
characteristics of this kind of data.

%% The Poisson distribution is indexed by a parameter $\lambda$ that is
%% its expectation and also its variance, i.e., its mean is equal to its
%% variance. This peculiarity of the Poisson model, imposes some
%% restrictions when regression models are built using this
%% distribution. The model imposes the assumption of equidispersion
%% (mean equals the variance) which may not be adequate for several
%% situations. If the Poisson model is applied to situations where the
%% equidispersion is not satisfied, the parameter estimates will be
%% inefficient and the standard errors will be
%% inconsistent \cite{Winkelmann1994}, \cite{Winkelmann1995}.

The assumption of variance equals to the mean (equidispersion)
underlying Poisson regression models imposes practical restrictions.
Parameter estimates will be inefficient, with inconsistent standard
errors,
%% and type I ou II error rates may increase when the Poisson model is
%% applied to non-equidispersed data.
and with larger error rates for hypothesis tests
%%, common after running a experiment, may increase
when the Poisson model is applied to non-equidispersed
data \cite{Winkelmann1994,Winkelmann1995}.

Overdispersion, with the variance greater than the mean, is largely
reported in the literature and may occur due to the absence of relevant
covariates, heterogeneity of sampling units, sampling levels, excess of
zeros \cite{Grunwald2011}. An usual approach is to adopt a generalized
linear mixed model (GLMM) describing the extra variability by the
inclusion of a non-observed latent random variable. An interesting case
is to assume a Poisson model with Gamma distributed random effects
leading to a negative binomial marginal distribution for the
responses. \citeasnoun{ElShaarawi2011} provides an overview of this and
other alternatives.

Lesser reported are the cases of underdispersion, with variances smaller
than the means. Explanatory mechanisms are more scarce and, typically,
heavily dependent on the context. A possible general description can be
derived by revisiting the key property of independent exponentially
distributed times between events underlying the Poisson model. If
%% the assumption of exponentially distributed times between events is 
inadequate, the occurrence of an event affects the probability of
another one, generating over or under dispersed counts. Other continuous
probability distributions with positive domain can be assumed such as
Gamma \cite{Winkelmann1995,Toft2006},
lognormal \cite{Gonzales-Barron2011} and
Weibull \cite{McShane2008}. Alternative approaches includes weighting
the Poisson distribution \cite{Ridout2004}, the COM-Poisson
distribution \cite{Lord2010,Lord2008} and heavy tail
distributions \cite{Zhu2009}.

%% We consider the analysis of an experiment designed to assess whether
%% and how the number of bolls in cotton plantas are affect by different
%% levels of defoliation during different phenological states
%% stages. Preliminary analysis indicates underdispersion which could
%% possibly be explained by competing for resources. A more detailed
%% analysis ....

\citeasnoun{Winkelmann1995} explores the connection between models for counts and models
for durations (lifetimes) relaxing the assumption of equidispersion at
the cost of an
%% extra parameter. The more general Gamma-count model assumes a Gamma
%% distribution for times
extra parameter denoted by $\alpha$. The Gamma-count model is a
convenient choice assuming Gamma distributed times between events. The
Poisson model becomes a particular case when the restriction $\alpha =
1$ implies the durations distribution reduces to the exponential
distribution. Varying values for the parameter $\alpha$ induces a
flexible probability distribution for the counts, which become
underdispersed for $\alpha > 1$ and overdispersed for $0 < \alpha < 1$.

We adopt the Gamma-count model for the analysis of a cotton production
agronomic experiment and compare the results against the ones obtained
with Poisson and quasi-Poisson models.
%% The experimental response variable is the number of cotton bolls
%% produced by each plant which is expected to be underdispersed given
%% the nutrient competition in the plant physiology.
Firstly, standard Poisson model is not excluded since it becomes a
particular case.
%% as the Poisson is a special case of Gamma-count model, considerar o
%% modelo Gamma-count não implica no abandono do modelo Poisson. when
%% holding the Gamma-count .
Secondly, fitting the Gamma-count model allows for investigating whether
the occurrences of bolls within a plant are independent events, an
arguable assumption under the simpler Poisson model.
%% usar o modelo Gamma-count permite verificar essa hipótese.
Thirdly, descriptive analyses of the data provided a clear empirical
evidence that the variance is a function of the mean with a constant of
proportionality below one.
%% If $\alpha = 1$, so Poisson model should be selected.
We also analyze the data by a semi-parametric quasi-Poisson model as the
benchmark for quantifying the observed variability in the data.

The Gamma-count regression model is not the canonical choice amongst
users of applied statistics and not widely available in statistical
software. For this reason, generic functions for maximum likelihood
inference are available as on-line supplements. This includes key
aspects related to inference upon the parameters of the Gamma-count
model, such as, construction of confidence intervals, either asymptotic
or based on profile likelihoods; hypothesis tests; model comparisons and
prediction with corresponding confidence intervals are also included,
all used throughout the data analysis.
%% once were not found in the literature for thi

\section{Background}\label{background}
Poisson regression models for count data follows directly from the
generalized linear model structure. Alternatively, the Poisson model can
be derived by assuming independent and exponentially distributed times
between events. The latter allows for the construction of alternatives
for under or overdispersed data such as the Gamma-count
model \cite{Winkelmann1995}, as follows below.

%% \citeasnoun{Winkelmann1995} discusses the relationship between the
%% times of occurrence and the counts leading to the Gamma-count model.
%% For sake of completeness the main theoretical aspects of Count-gamma
%% model developed by \citeasnoun{Winkelmann1995} will be reported here.

Elementary probability arguments establish that the distribution of a
count variable can be derived from the distribution of arrival times.
Let $\tau_k > 0$, $k \in \mathbb{N}$, denote a sequence of waiting times
between the ($k-1$) and the $k^{th}$ event. Then, the arrival time of
the $n^{th}$ event is
\begin{equation}\label{eq:varthetan}
 \vartheta_n = \sum_{k=1}^{n} \tau_k, \qquad n = 1, 2, \ldots .
\end{equation}
Let $N_T$ represent the total number of events within a $(0, T)$
interval. $N_T$ is a count variable. It follows from the definition of
$N_T$ and $\vartheta_n$ that
\begin{align}
\nonumber N_T < n \quad &\iff \quad \vartheta_n \geq T \\
\nonumber \Pr(N_T < n) &= \Pr(\vartheta_n \geq T) = 1-F_n(T) \\
 \Pr(N_T=n) &= F_n(T) - F_{n+1}(T), \label{eq:n}
\end{align}
%% \[
%% N_T < n \quad \iff \quad \vartheta_n \geq T.
%% \]
%% As a consequence
%% \[
%% \Pr(N_T < n) = \Pr(\vartheta_n \geq T) = 1-F_n(T),
%% \]
where $F_n(T)$ is the cumulative distribution function of $\vartheta_n$. 
%% Furthermore,
%% \begin{equation}\label{eq:n}
%% \Pr(N_T=n) = F_n(T) - F_{n+1}(T).
%% \end{equation}
%% Equation (\ref{eq:n}) provides the fundamental relation between the
%% distribution of arrival times and the distribution of counts. The
%% distribution of $N_T$ can be obtained from knowledge of the distribution
%% of $\vartheta_n$.
Equation (\ref{eq:n}) allows obtaining the distribution of counts $N_T$
from knowledge of the distribution of arrival times $\vartheta_n$.

%% It will be assumed that $\tau_k$ are identically and independently
%% Gamma distributed. Dropping the index $k$ the density can be written
%% as

It is assumed $\tau_k$ are identically and independently Gamma
($G(\alpha, \beta)$) distributed with density:
\begin{equation*}
 f(\tau; \alpha, \beta) = \frac{\beta^\alpha}{\Gamma(\alpha)}\,\,
 \tau^{\alpha-1} \,\, \exp\{-\beta\tau\}, \qquad \alpha, \beta \in \mathbb{R}^{+}.
\end{equation*}
with $\tau >0$, mean $\text{E}(\tau) = \alpha/\beta$ and variance
$\text{Var}(\tau) = \alpha/\beta^2$. By (\ref{eq:varthetan}),
$\vartheta$ is the sum of iid Gamma random variables therefore with
density $G(n\alpha, \beta)$.
%% By properties of gamma distribution the density function of
%% $\vartheta_n$ is
%% \begin{equation}
%% f_n(\vartheta; \alpha, \beta) = \frac{\beta^{n\alpha}}{\Gamma(n\alpha)}\,\,
%% \tau^{n\alpha-1} \,\, \exp\{-\beta\vartheta\}.
%% \end{equation}
%% To derive the count distribution, we have to evaluate the cumulative
%% distribution function
Let $G(n\alpha,\beta T)$ be the cumulative distribution function
evaluated at $\beta T$:
\begin{equation}\label{eq:G}
 G(n\alpha,\beta T) = F_n(T) = \frac{1}{\Gamma(n\alpha)}
\int_{0}^{\beta T} u^{n\alpha-1} \,\, \exp\{-u\} \,\, \text{d}u.
\end{equation}
The count distribution~(\ref{eq:n}) for number of events within the time
interval (0,$T$) is given by:
\begin{equation}\label{eq:ndens}
 \Pr(N=n) = G(\alpha n, \beta T)- G(\alpha(n+1), \beta T) ,
\end{equation}
with expected value given by:
\begin{equation}\label{eq:nexpected}
 \text{E}(N_T) = \sum_{i=1}^{\infty} G(\alpha i, \beta T).
\end{equation}
For $\alpha=1$, $f(\tau)$ reduces to the exponential density and
(\ref{eq:ndens}) simplifies to the Poisson distribution.

For the Gamma-count regression model the parameters depend on a vector
of individual covariates, indicated by the subscript $i$. Assuming that
the period at risk is the same for all observation, $T$ can be set to
unity, without loss of generality.
%% Assume that
%% \begin{equation}
%% \frac{\beta}{\alpha} = \exp\{x_i^\top \gamma\}.
%% \end{equation}
This yields the regression
\begin{equation*}
 \text{E}(\tau_i|x_i) = \frac{\alpha}{\beta} = \exp\{-x_i^\top \gamma\}.
\end{equation*}
Is important do emphasize that the regression is for the waiting times
$\tau_i$ and not for the counts $N_i$ since $\text{E}(N_i|x_i) =
(\text{E}(\tau_i|x_i))^{-1}$ does not holds unless $\alpha=1$. For a
given $\gamma$, $\text{E}(N_i|x_i)$ is evaluated
by~(\ref{eq:nexpected}).

%% Consider that the times between events are independent, but no longer
%% exponentialy distributed, rather follow some other distribution with
%% a non-constant hazard function, such as a gamma distribution, as
%% summarised in Figure \ref{fig:modelo}.

\begin{figure}[!th]
\begin{center}
\includegraphics[width=0.98\textwidth]{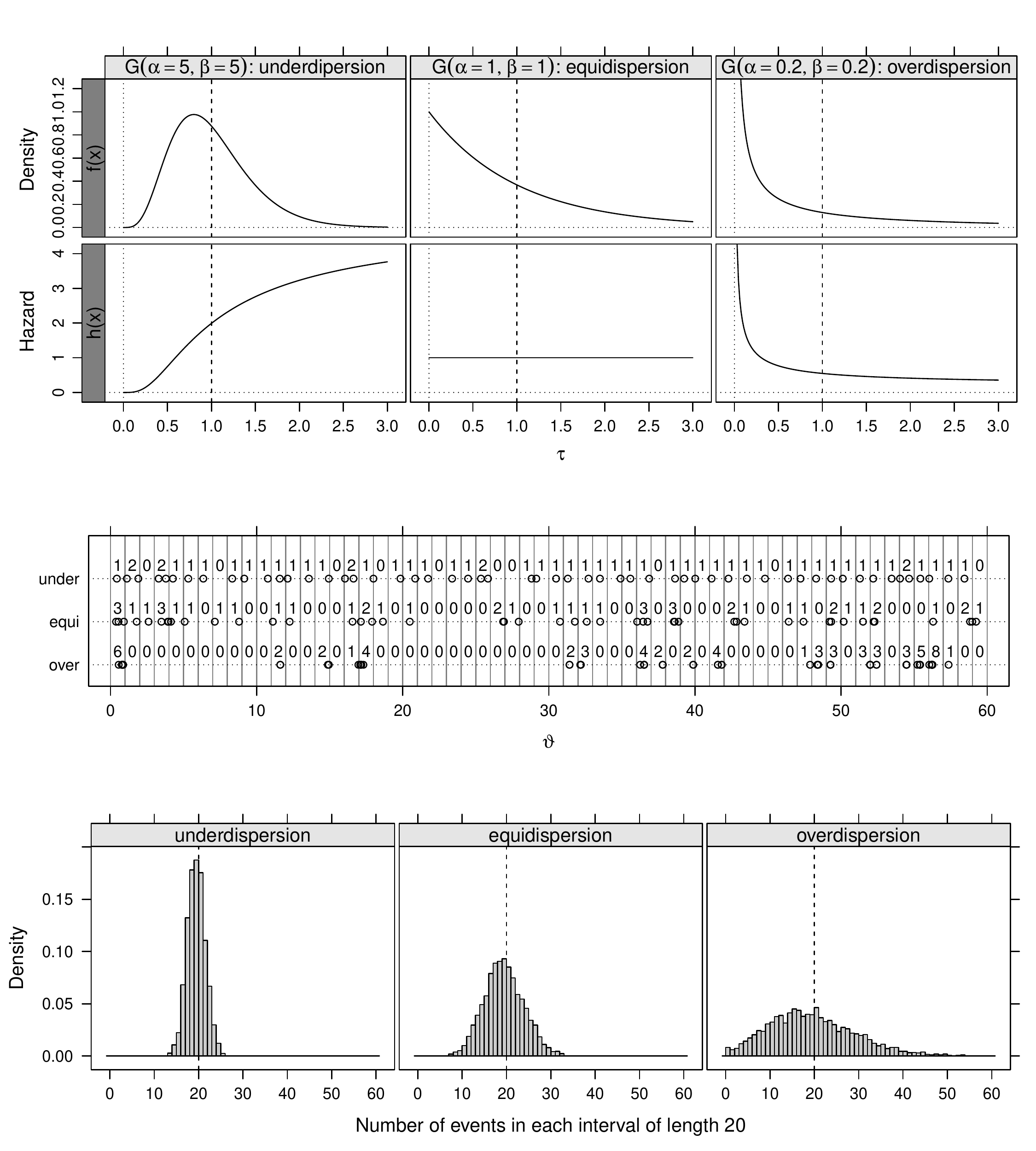}
\end{center}
\caption{Comparison of different distribution of time between events. Top panel:
Gamma densities and hazard functions, middle panel: simulated events and
corresponding interval counts for each distribution, bottom panel:
counts histograms.}\label{fig:modelo}
%% \caption{Count patterns according for different distribution of time
%% between events: (upper top) Gamma densities $f(x)$, (middle top)
%% hazard functions $h(x)$, (middle botton) number of events per each
%% unit interval and (lower botton) frequency distribution of number of
%% events per interval of length 20.}\label{fig:modelo}
\end{figure}

Figure~\ref{fig:modelo} illustrates the relation between the
distribution of times between events and counts showing the graphics of
density and hazard functions with corresponding simulated values. Gamma
distributions with unity mean and different variances are shown in the
first line. The second line displays the corresponding increasing,
constant and decreasing hazard functions related to smaller, equal or
larger variances than the mean. The middle plots correspond to the
exponential distribution and its constant hazard function.
%% which implies on the memoryless property of an exponential random
%% variable.
The middle panel show simulated values with time intervals drawn from
each of the above mentioned distributions. Vertical lines indicates
fixed width intervals for which events are counted and the counts within
each interval are displayed.
%% Simulated values of $\tau$ from these distributions where obtained
%% and organized in a line with fixed marked intervals. The number,
%% $N$, of values in each interval is displayed.
The distribution of events is nearly regular and the counts have smaller
variance in the underdispersed case. For the overdispersed case, the
events are clustered with large variances for counts. Differences are
evident in the resulting histograms.

%% caso m s2 cv
%% 1 over 1.0333333 3.1514124 304.97540
%% 2 equi 0.7833333 0.7827684 99.92788
%% 3 under 0.9166667 0.2471751 26.96456
%% \begin{table}[ht]
%% \centering
%% \begin{tabular}{rlrrr}
%% \hline
%% & caso & m & s2 & cv \\ 
%% \hline
%% 1 & over & 1.03 & 3.15 & 304.98 \\ 
%% 2 & equi & 0.78 & 0.78 & 99.93 \\ 
%% 3 & under & 0.92 & 0.25 & 26.96 \\ 
%% \hline
%% \end{tabular}
%% \end{table}

%% The distribution between the events is Gamma for the three cases,
%% with the same mean but different variances. The variance for the
%% first is $5/5^2$, consequently the hazard function is increasing,
%% generating underdispersed counts. For the case with largest variance
%% ($0.2/0.2^2$) the hazard function is decreasing generating
%% overdispersed counts. For the canonical situation with variance
%% equals $1$, the Gamma distribution reduces to an exponential
%% distribution and the hazard function is constant generating
%% equidispersed counts. Figure~\ref{fig:modelo} illustrates the
%% limitation of the Poisson distribution which would be suitable for
%% modelling counts in just one out of the tree possibilities.

For a sample if independent counts $y_i, i=1 \ldots n$, estimates
$\hat{\alpha}$ and $\hat{\gamma}$ can be obtained by maximizing the
log-likelihood
\begin{equation}\label{eq:vero}
\ell(\gamma, \alpha; y, x) = \sum_{i=1}^n \log \left(
G(y_i \alpha, \alpha \exp(x_i^{\top} \gamma)) -
G(y_i(\alpha+1), \alpha
\exp(x_i^{\top} \gamma)) \right), 
\end{equation}
where $\gamma$ is the vector of regression parameters describing the
interval between the events, $\alpha$ is the dispersion parameter, $x_i$
is a vector of covariates and $G()$ is given by~(\ref{eq:G}).
%% The particular case of the Poisson model with $\alpha = 1$ results
%% from the difference between two cumulative exponential density
%% functions.

Parameter estimation requires numerical maximization of~(\ref{eq:vero}).
Confidence intervals and hypotheses tests can be either based upon
quadratic approximations of the likelihood function (Wald type
intervals) or profile likelihoods.

%% Predicted values and confidence bands are obtained as follows. By
%% the delta method.
For a vector $x$ of covariates values, time between events is predicted
by:
\begin{equation*}
 \hat{\eta} = x^\top \hat{\gamma}.
\end{equation*}
The covariance matrix for the model parameters is:
\begin{equation*}
V = \begin{bmatrix}
V_{\alpha\alpha} \quad V_{\alpha\gamma} \\
V_{\gamma\alpha} \quad V_{\gamma\gamma}
\end{bmatrix},
\end{equation*}
and estimated by the negative of the inverse Hessian matrix numerically
obtained around the maximised log-likelihood.
%% Let $j$ denote the set of indices regarding nuisance parameters
%% ($\alpha)$ and $-j$ that parameters in linear predictor ($\gamma$).
%% In this case $\alpha$ is a nuisance parameter.
The prediction standard error is given by:
\begin{equation*}
 \text{se}(\hat{\eta}) = \sqrt{x^\top V_{\gamma|\alpha}\, x},
\end{equation*}
where $V_{\gamma|\alpha} = V_{\gamma\gamma}-V_{\gamma\alpha}
V_{\alpha\alpha}^{-1} V_{\alpha\gamma}$. For the paticular case of the
Gamma-count model considered here $\alpha$ is an scalar and found to be
nearly orthogonal to $\gamma$ in which case $V_{\gamma|\alpha} \approx
V_{\gamma\gamma}$.
%% Confidence intervals are obtained at the scale of the linear
%% predictor.
Confidence intervals for the mean counts are obtained by
computing~(\ref{eq:nexpected}) after transform the limits of the
confidence interval on the scale of the linear predictor by the inverse
link function $g^{-1}()$.

%% Let $z_{\delta/2}$ a quantile of standard normal for a $1-\delta$
%% nominal coverage interval. The confidence interval for a predicted
%% value is given by
%% \begin{equation}
%% (\hat{\eta}_l; \hat{\eta}_u) = \hat{\eta}\pm z_{\alpha/2}\, \text{se}(\hat{\tau}).
%% \end{equation}
%% Now, each value ($\hat{\eta}_l$, $\hat{\eta}$ and $\hat{\eta}_u$) are
%% used in equation (\ref{eq:nexpected}), where $\beta = \exp\{\eta\}$,
%% to get confidence bands for the count variable $N$.

\section{Data set and models}\label{dataset}
\begin{figure}[!th]
\begin{center}
 \includegraphics[width=0.95\textwidth]{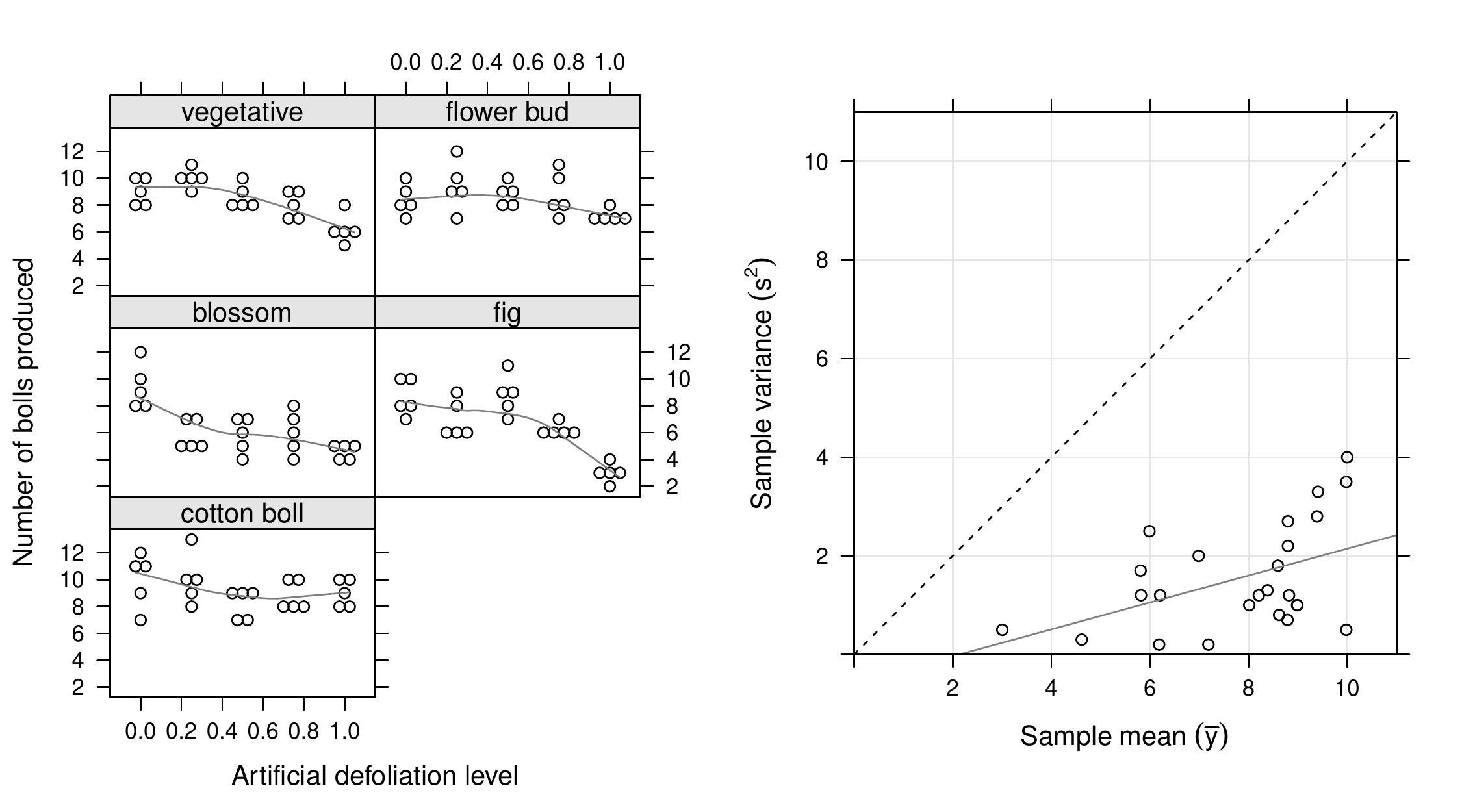} 
\end{center}
\caption{
(left) Number of bolls produced for each artificial defoliation level
and each growth stage. (right) Sample variance against the sample mean
of the five replicates for each combination of defoliation level and
growth stage.}\label{fig:descritiva}
\end{figure}

The data that motivates this paper come from a greenhouse experiment
with cotton plants (\textit{Gossypium hirsutum}) obtained under a
completely randomized design with five replicates. The experiment aimed
to assess the effects of five defoliation levels $(0,25,50,75,100\%)$ on
the observed the number of bolls produced by plants at five growth
stages: vegetative, flower-bud, blossom, fig and cotton
boll \cite{Silva2012}. The experimental unity was a vase with two
plants. The number of cotton bolls was recorded at the each culture
cycle. Figure~\ref{fig:descritiva} (left) shows the number of cotton
bolls recorded for each combination of defoliation level and growth
stage. All the points in the sample means and variances dispersion
diagram (right) are below the identity line, clearly suggesting the
presence of underdispersion.

The analysis and assessment of the effects of the experimental factors
are based on the Gamma-count, Poisson and quasi-Poisson models, with the
following structures for the log-link function $g()$:
%% a sequence starting from the single intercept up to the model with
%% quadratic effect of defoliation for each growth stage. The analysis
%% consists of specifying a sequence of five models adding terms to the
%% model in order to allow for testing significance of the experimental
%% factors. Each model was proposed in order to test different
%% structures of the effects in the experiment. The structure of the
%% considered models for the link function $g()$ are the following:
\begin{enumerate}
\item[] Predictor 1: $g(\mu) = \gamma_0 $;
\item[] Predictor 2:
  $g(\mu) = \gamma_0 + \gamma_1 \text{def}$ (first order effect of
  defoliation);
\item[] Predictor 3:
  $g(\mu) = \gamma_0 + \gamma_1 \text{def} + \gamma_2 \text{def}^2$
 (second order effect of defoliation);
\item[] Predictor 4:
  $g(\mu) = \gamma_0 + \gamma_{1j} \text{def} + \gamma_2 \text{def}^2$
 (first order defoliation effect for each growth stage);
\item[] Predictor 5:
  $g(\mu) = \gamma_0 + \gamma_{1j} \text{def}
 + \gamma_{2j} \text{def}^2$ (second order effect defoliation for each
 growth stage).
\end{enumerate}

The parameter $\mu$ is the expected value of $N$ for the Poisson and
quasi-Poisson models and the expected value of the latent random
variable $\tau$ equivalent to \textit{time between events} for the
Gamma-count model.

%% The nested structure of the predictors eases the conduction of
%% hypothesis tests based on likelihood ratios.
The nested structure of the predictors allows relevant hypothesis tested
by likelihood ratios. Predictor~1 contains only the intercept and is
fitted simply as a baseline to assess to which extent the structured
models improve the fit. Linear and quadratic effects of defoliation are
added by Predictor~2 and Predictor~3, respectively. Predictor~4 and
Predictor~5 allows the linear and quadratic effects of defoliation to
vary between the growth stages, as indicated by the subscript $j$. The
parameter $\gamma_0$ is not allowed to vary between the growth stage
once the effect of no defoliation is the same for all growth stages.

Values of the maximized log-likelihood and the Akaike criterion are
recorded for the fully parametric Poisson and Gamma-count models. The
semi-parametric quasi-Poisson model is also fitted to assess whether the
parametric models produce comparable results. This model is less
restrictive concerning model assumptions, albeit without the inferential
advantages of the fully parametric counterparts.
%% Poisson and gamma-count .

\section{Results}\label{results}
Table~\ref{lrt} summarises the maximised log-likelihoods and likelihood
ratio tests comparing the sequence of predictors for the Poisson and
Gamma-count models, as well as fitting results for the
quasi-Poisson. The Gamma-count model has a higher log-likelihood
%% when compared to the Poisson 
with the hypothesis of equidispersion ($\alpha=1$) being rejected by
likelihood ratio tests, even for the predictor without covariates.
Estimates of $\hat\alpha > 1$ confirms the number of cotton bolls are
underdispersed with
%% , positive dependence duration. 
increasing hazard functions indicating that the probability of the
development of a new cotton boll increases as
\textit{time} progresses. This result supports the hypothesis of a
regular sharing of plant resources in the distribution of the number of
cotton bolls.
%% In other words, resources should be acumulated at a certain level
%% $\tau$ to produce a boll. To produce at least $N$ bolls the plant
%% need acumulated $\vartheta_n = \sum \tau$.
The quasi-Poisson model also indicates underdispersion ($\phi<1$), even
for the null model.

\begin{table}[t]
\caption{
Model fit measures and comparisons between predictors and
models.}\label{lrt}
% Pré-visualizar código-fonte para parágrafo 0

\begin{tabular}{rrrrrrrrr}
\hline 
Poisson & np & $\ell$ & AIC & diff np  & 2(diff $\ell$) & P($>\chi^2$) &  & \tabularnewline
\hline
1 & 1 & -279.933 & 561.866 &  &  &  &  & \tabularnewline
2 & 2 & -272.001 & 548.001 & 1 & 15.864 & 6.805E-05 &  & \tabularnewline
3 & 3 & -271.354 & 548.709 & 1 & 1.293 & 2.556E-01 &  & \tabularnewline
4 & 7 & -258.674 & 531.348 & 4 & 25.360 & 4.258E-05 &  & \tabularnewline
5 & 11 & -255.803 & 533.606 & 4 & 5.742 & 2.193E-01 &  & \tabularnewline
\hline
\hline 
Gamma-count & np & $\ell$ & AIC & diff np & 2(diff $\ell$) & P($>\chi^2$) & $\hat\alpha$ &
P($>\chi^2$)$^a$\tabularnewline
\hline
1 & 2 & -272.396 & 548.792 &  &  &  & 1.764 & 1.034E-04\tabularnewline
2 & 3 & -257.350 & 520.701 & 1 & 30.092 & 4.121E-08 & 2.266 & 6.198E-08\tabularnewline
3 & 4 & -255.981 & 519.962 & 1 & 2.738 & 9.796E-02 & 2.317 & 2.940E-08\tabularnewline
4 & 8 & -220.145 & 456.291 & 4 & 71.671 & 1.007E-14 & 4.206 & 1.661E-18\tabularnewline
5 & 12 & -208.386 & 440.773 & 4 & 23.518 & 9.976E-05 & 5.112 & 2.071E-22\tabularnewline
\hline
\hline 
Quasi-Poisson & np & deviance &  & diff np & diff dev & P($>F$) & $\hat\phi$ &
P($>\chi^2$)$^a$\tabularnewline
\hline
1 & 1 & 75.514 &  &  &  &  & 0.567 & 3.660E-04\tabularnewline
2 & 2 & 59.650 &  & 1 & 34.214 & 4.235E-08 & 0.464 & 5.134E-07\tabularnewline
3 & 3 & 58.357 &  & 1 & 2.810 & 9.630E-02 & 0.460 & 3.661E-07\tabularnewline
4 & 7 & 32.997 &  & 4 & 22.768 & 7.676E-14 & 0.278 & 9.154E-16\tabularnewline
5 & 11 & 27.255 &  & 4 & 5.956 & 2.241E-04 & 0.241 & 3.566E-18\tabularnewline
\hline
\end{tabular}

{\small np - number of parameters; $\ell$ - log-likelihood; diff np -
difference in np; diff $\ell$ - difference in $\ell$; diff dev -
difference in scaled deviance; $^a$bilateral hypothesis test of
dispersion parameter equal to 1.}
\end{table}

Unlike the others, the Poisson model does not shown significant effects
under Model~5.
%% This can be attributed to the inadequate assumption of equidispersion.
This is attributed to the inadequate assumption of equidispersion that
makes the log-likelihood among predictors less distinguishable.
Descriptive levels ($p$-values) are substantially smaller for the
Gamma-count and quasi-Poisson, compared with the Poisson model. In the
presence of underdispersion the latter becomes conservative for
hypothesis testing.
%% Although these figures do not allow assessing the model fitting it is
%% clear that in the presence of underdispersion the Poisson model
%% becomes conservative for hypothesis tests.

\begin{figure}[!th]
\begin{center}
 \includegraphics[width=0.95\textwidth]{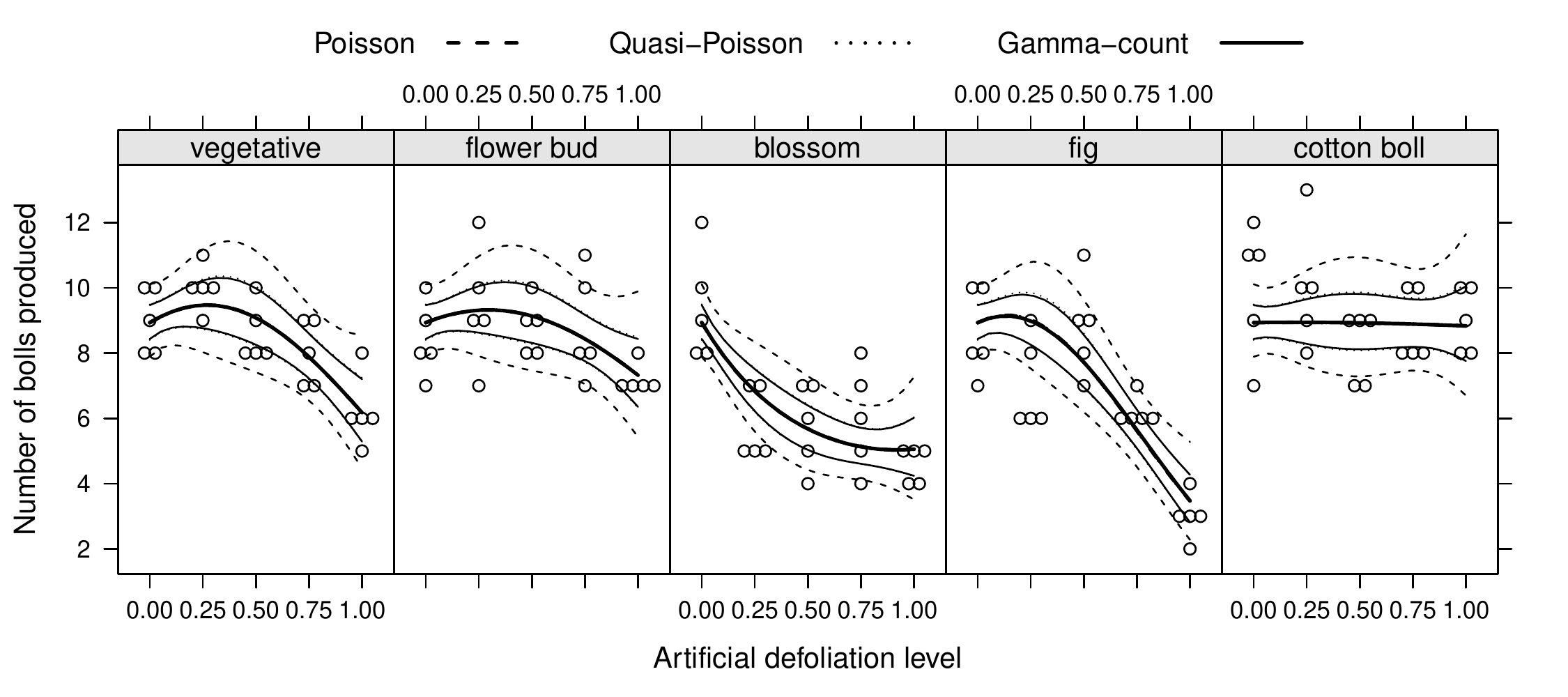} 
\end{center}
\caption{
Dispersion diagrams of observed values and curves of predicted values
and confidence intervals (95\%) as functions of defoliation level for
each growth stage.}\label{bandas}
\end{figure}

The Gamma-count and the quasi-Poisson models indicate that both, linear
and quadratic effects of levels of defoliation, vary between growth
stages. Results on Table~\ref{estimates} and Figure~\ref{bandas} show,
for all models, no significant effects of defoliation during the
floral-bud and cotton boll stages. The ratio between the estimates and
the corresponding standard errors for these stages are, in absolute
values, smaller than the reference value of 1.96 for a significance
level of 5\%. The Poisson model only detects the effect of defoliation
for the blossom stage, while the Gamma-count and quasi-Poisson models
indicate a significant effect of defoliation for the vegetative, blossom
and fig stages.

\begin{table}[!th]
\caption{
Parameter estimates and estimate/standard error rates for the three
models.}\label{estimates}
%% \caption{Parameter estimates and ratio estimate/standard error by
%% three models all with logarithm link function.}\label{estimates}
% Pré-visualizar código-fonte para parágrafo 1

\begin{tabular}{lrrrrrrrr}
\hline 
 & \multicolumn{2}{c}{Poisson} &  & \multicolumn{2}{c}{quasi-Poisson} &  & \multicolumn{2}{c}{Gamma-count}\tabularnewline
\cline{2-3} \cline{5-6} \cline{8-9} 
Parameter & Estimate  & Est/SE  &  & Estimate  & Est/SE  &  & Estimate  & Est/SE\tabularnewline
\hline
$\gamma_0$  		& 2.1896 & 34.5724* &  & 2.1896 & 70.4205* &  & 2.2342 &
79.7128*\tabularnewline
$\gamma_{1vegetative}$  	& 0.4369 & 0.8473 &  & 0.4369 & 1.7260 &  & 0.4122 &
1.8080\tabularnewline
$\gamma_{2vegetative}$  	& -0.8052 & -1.3790 &  & -0.8052 & -2.8089* &  & -0.7628 &
-2.9544*\tabularnewline
$\gamma_{1bud}$  	& 0.2897 & 0.5706 &  & 0.2897 & 1.1622 &  & 0.2744 & 1.2224\tabularnewline
$\gamma_{2bud}$  	& -0.4879 & -0.8613 &  & -0.4879 & -1.7544 &  & -0.4642 &
-1.8534\tabularnewline
$\gamma_{1blossom}$  	& -1.2425 & -2.0581* &  & -1.2425 & -4.1921* &  & -1.1821 &
-4.4348*\tabularnewline
$\gamma_{2blossom}$  	& 0.6728 & 0.9892 &  & 0.6728 & 2.0149* &  & 0.6453 & 2.1486*\tabularnewline
$\gamma_{1fig}$  	& 0.3649 & 0.6449 &  & 0.3649 & 1.3135 &  & 0.3198 & 1.2797\tabularnewline
$\gamma_{2fig}$  	& -1.3103 & -1.9477 &  & -1.3103 & -3.9672* &  & -1.1990 &
-4.0385*\tabularnewline
$\gamma_{1boll}$  	& 0.0089 & 0.0178 &  & 0.0089 & 0.0362 &  & 0.0070 & 0.0315\tabularnewline
$\gamma_{2boll}$  	& -0.0200 & -0.0361 &  & -0.0200 & -0.0736 &  & -0.0185 &
-0.0756\tabularnewline
$\alpha$  & - & - &  & - & - &  & 5.1120 & 7.4228*\tabularnewline
\hline
\end{tabular}
 \\
{\small *indicates $|\mathrm{Est/SE}| > 1.96$. }
\end{table}

Parameter estimates for the blossom stage have opposite signal when
compared to the other stages. A negative and significant linear term
indicates a rapid decay in the number of cotton bolls during the
beginning of defoliation. The positive quadratic term indicates concave
up response as seen in Figure \ref{bandas} for the blossom
stage. Therefore, the impact of defoliation is greater for the blossom
stage and there is a tolerance up to approximately 40\% of defoliation
for the vegetative stage and 24\% for the fig stages.
%% The precision of estimates are, on average, of 3\% in order of
%% magnitude were higher for the Gamma-count model.
Parameter estimates between models are not directly comparable once they
are related to the number of events in the Poisson model and to the
distribution of the time between events for the Gamma-count model.

%% In the Gamma-count model, the estimated parameters are related , not
%% to the number of events such as the Poisson and quasi-Poisson
%% models. Despite not being directly observed, the distribution of time
%% between events can be estimated from the observed number of events
%% under the Gamma-count model. This is an interesting feature of the
%% model, since the number of events is typically easier to register.

Prediction curves for each stage are shown in Figure~\ref{bandas} and
are indistinguishable between the three models. The confidence bands are
similar between Gamma-count and quasi-Poisson models and clearly wider
for the Poisson model.

\begin{figure}[!th]
\begin{center}
 \includegraphics[width=0.5\textwidth]{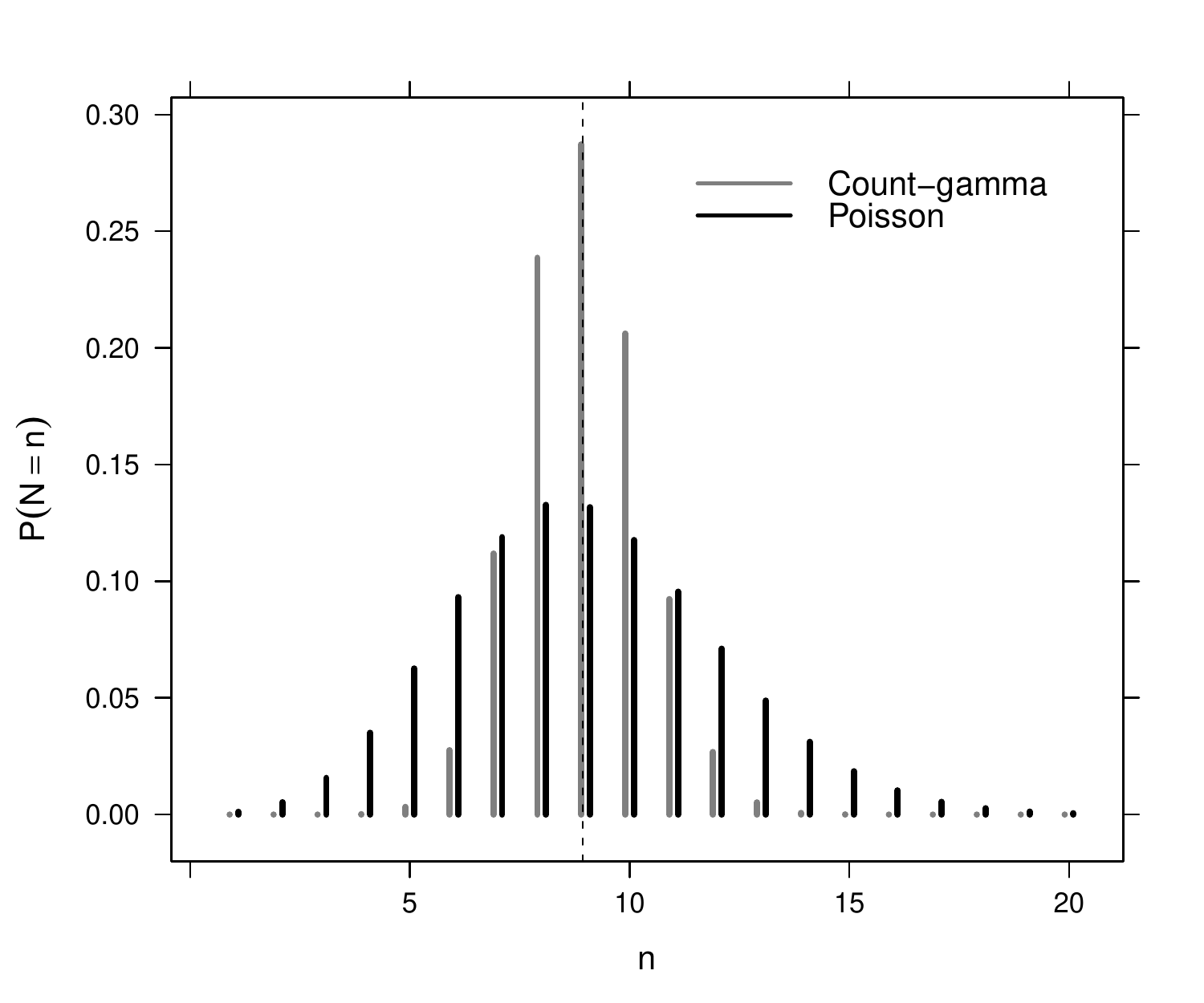} 
\end{center}
\caption{
Estimated probabilities from Poisson and Gamma-count models for a level
zero of defoliation.}\label{prob}
\end{figure}

Overall the Gamma-count and the quasi-Poisson model produced very
similar inferential results, point and interval estimates, hypothesis
tests, model comparisons and prediction bands.
%% This shows that the Gamma-count model is highly flexible. 
The semi-parametric quasi-Poisson model is expected to have a better fit
to a particular data set, as there is no explicit formulation of a
probability model and functional relation between mean and
variance. Such flexibility comes with drawbacks. There are no likelihood
measures for comparing models and submodels neither an estimated
probability distribution for the counts, which could address questions
of scientific interest. Table~\ref{prob} provides the estimated
probability distributions for the number of cotton bolls obtained under
Poisson and Gamma-count models. At the level zero of defoliation, the
expected value is 8.93 cotton bolls per two plants for either model,
however with probability distribution more concentrated around the mean
value under the Gamma-count model.

\begin{figure}[!th]
\begin{center}
 \includegraphics[width=0.45\textwidth]{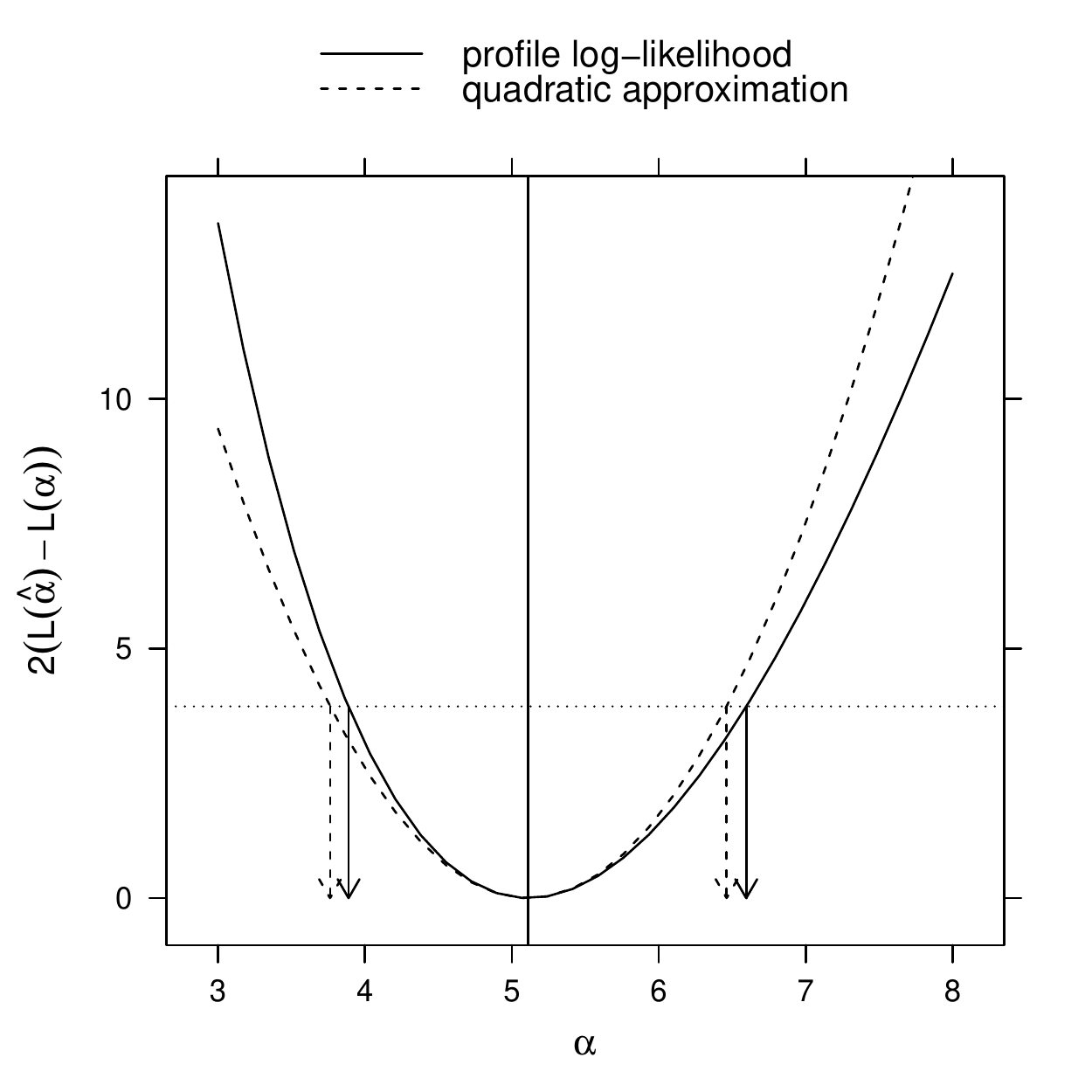}
 \includegraphics[width=0.45\textwidth]{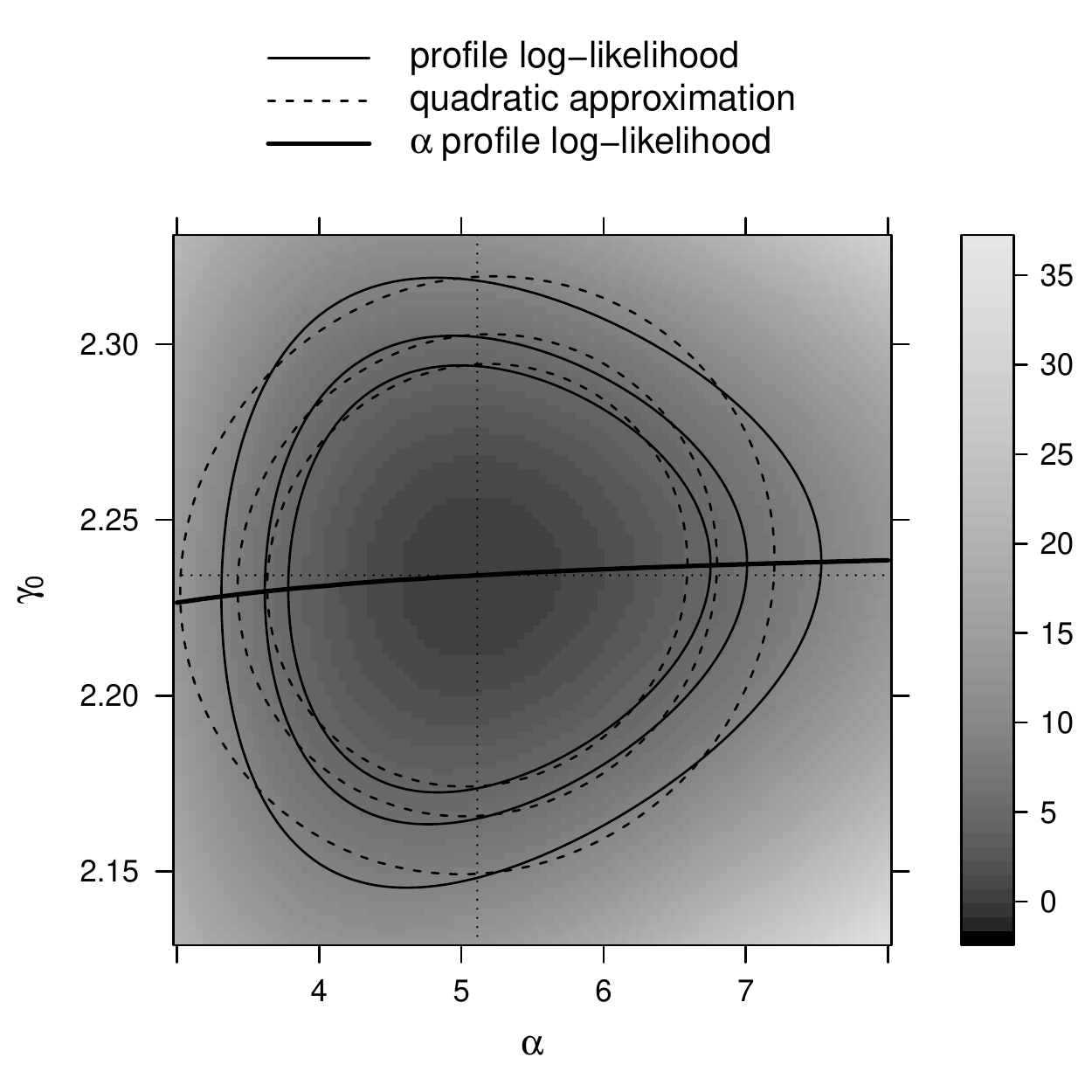} 
\end{center}
\caption{
Profile likelihood and quadratic approximation for: (left) $\alpha$ with
arrows indicating the 95\% confidence intervals and (right) $(90, 95,
99\%)$ confidence regions for $\gamma_0$ and $\alpha$.}\label{ic}
\end{figure}

%% Despite the great potential for usage, the Gamma-count has not being
%% considered and assessed in agronomic studies.
In what follows we further explore aspects of the likelihood function.
The profile log-likelihood for $\alpha$ is slightly skewed (left panel,
Figure~\ref{ic}). The 95\% confidence interval based on the $\chi^2$
distribution is $(3.89, 6.59)$ while the asymptotic interval is $(3.76,
6.46)$. Both have the same range (2.70) however shifted by 0.13
units. This is a small difference and the quadratic approximation of the
likelihood is considered satisfactory. Although the precision of the
intervals are similar, the interval based on the log-likelihood is
preferred to describe the uncertainty associated with $\alpha$ since it
is able to detect possible asymmetries and has limits within the
$(0, \infty)$ parameter space.

The right panel in Figure~\ref{ic} shows the confidence regions for
$\alpha$ and $\gamma_0$ obtained via profile likelihood and quadratic
approximation of the likelihood. Axes of the confidence regions are
nearly parallel to the Cartesian axes suggesting the parameters are
nearly orthogonal. Moreover, covariances between $\hat\alpha$ and each
of the other parameters $\hat\gamma$ (not shown) are nearly zero
implying the inferences about one parameter are not influenced by the
other parameter.
%% This is an useful property implying inferences about a parameter are
%% not strongly influenced by other parameters.
The confidence regions are symmetric in the direction of $\gamma_0$ and
the asymptotic and profile likelihood based confidence intervals are
therefore coincident.

Computationally, the asymptotic confidence interval is easier to obtain
since it simply requires the inversion of the Hessian matrix around the
maximum of the log-likelihood function. The profile log-likelihood
requires successive optimizations for a set of values of the parameter
of interest. For a larger number of parameters obtaining individual
intervals based on the profile the likelihoods will increase the
computational burden.

\section{Conclusion}\label{conclusion}
The Poisson, Gamma-count and semi-parametric quasi-Poisson models were
considered for the analysis of underdispersed count responses from a
greenhouse experiment with cotton plants subjected to different
artificial defoliation levels and growth stages.

Significance of experimental factors are the same for the Gamma-count
and quasi-Poisson models whereas the Poisson model is more conservative,
not identifying some experimental factors as significant. The latter
have led to greater standard errors and wider prediction bands, being
unable to capture information contained in the data. The analysis
suggest that, in the presence of underdispersion, the standard Poisson
model is inadequate and can lead to wrong conclusions about the effects
of experimental factors or covariates of interest.

Results under the Gamma-count model are comparable to the
semi-parametric approach which does not assume an specific probability
distribution for the counts. The fully parametric approach is
advantageous since it allows for likelihood based inference, deriving
estimated prediction probabilities besides enabling generalizations such
as specifying a regression model structure also for the dispersion
parameter.

Likelihood analysis showed nearly quadratic behavior for the parameter
$\alpha$ controlling the dispersion of the counts. This parameter has
little influence upon point estimates of the regression parameters,
being responsible for stabilizing the estimates of variances of
regression parameters, which are often overestimated under the Poisson
distribution.

%% Despite the great potential for usage, the Gamma-count has not being
%% considered and assessed in agronomic studies.
Despite the advantages and potential for usage, the Gamma-count model is
uncommon
%% in the anaylsis of agronomical experiements and provides a
relevant addition to the suite of models to be considered for the
analysis of experimental count data. The model can be easily implemented
in a statistical programming language as illustrated by the
supplementary material.

Possible topics for further investigation and extensions include
assessment of impacts of misspecification under different levels
dispersion, increase of flexibility possibly by modeling the dispersion
parameter as function of covariates and the addition of random effects
to account for grouped data structures such as repeated and longitudinal
measures.

\bibliographystyle{ECA_jasa}
\bibliography{underdispersed}

\end{document}